\documentclass{aa}
\usepackage{graphics}
\usepackage{natbib}
\bibpunct{(}{)}{;}{a}{}{,}

\newcommand{\Lya}{Lyman-$\alpha$}
\newcommand{\Lyb}{Lyman-$\beta$}
\newcommand{\cq}{\ion{C}{iv}}

\newcommand{\cdue}{\ion{C}{ii}}

\newcommand{\siq}{\ion{Si}{iv}}
\newcommand{\sitre}{\ion{Si}{iii}}
\newcommand{\sidue}{\ion{Si}{ii}}
\newcommand{\mgd}{\ion{Mg}{ii}}

\newcommand{\fetre}{\ion{Fe}{iii}}
\newcommand{\fed}{\ion{Fe}{ii}}
\newcommand{\nc}{\ion{N}{v}}

\newcommand{\nuno}{\ion{N}{i}}
\newcommand{\huno}{\ion{H}{i}}
\newcommand{\ouno}{\ion{O}{i}}
\newcommand{\osei}{\ion{O}{vi}}
\newcommand{\aldue}{\ion{Al}{ii}}
\newcommand{\altre}{\ion{Al}{iii}}

\newcommand{\sdue}{\ion{S}{ii}}
\newcommand{\cm}{cm$^{-2}$}
\newcommand{\kms}{km s$^{-1}$}
\newcommand{\lsim}{\raisebox{-5pt}{$\;\stackrel{\textstyle <}{\sim}\;$}}
\newcommand{\gsim}{\raisebox{-5pt}{$\;\stackrel{\textstyle >}{\sim}\;$}}

\begin{document}


   \title{
High matter density peaks from UVES observations of
   QSO pairs: correlation properties and chemical
   abundances\thanks{Based on
   material collected with the European Southern
   Observatory Very Large Telescope operated on Cerro Paranal
          (Chile). Proposals 65.O-0299 and 67.A-0078}}

   \author{V. D'Odorico\inst{1} \and
   P. Petitjean\inst{1,2} \and S. Cristiani\inst{3,4}}

   \institute{Institut d'Astrophysique de Paris, 98bis Boulevard
        Arago, F-75014 Paris  
     \and
        LERMA, Observatoire de Paris, 61 Av. de
        l'Observatoire, F-75014 Paris
     \and
	European Southern Observatory,
        Karl-Schwarzschild-Strasse 2, D-85748 Garching,
        Germany
     \and
	Osservatorio Astronomico di Trieste, via
        G.B. Tiepolo, 11, I-34131 Trieste, Italy}
		
   \offprints{V. D'Odorico}
%
 
   \date{Received; accepted }

    \titlerunning{High matter density peaks from UVES observations of
   QSO pairs}
 
    \authorrunning{}

\abstract{We study the transverse clustering properties
      of high matter density peaks as traced by high
      column density absorption systems (either Lyman
      limit systems characterized by 
      $N($\huno$) \ge 2 \times 10^{17}$ \cm\ or \cq\
      systems with $W_{\rm r} > 0.5$ \AA) at redshifts
      between 2 and 3 with UVES spectra of two QSO pairs
      (UM680/UM681 at 56 arcsec angular separation and
      Q2344+1228/Q2343+1232 at 5 arcmin angular
      separation) and a QSO triplet
      (Q2139-4433/Q2139-4434/Q2138-4427 
      at 1, 7 and 8 arcmin angular separation). We find 3  
      damped \Lya\ systems ($N($\huno$) \ge 2
      \times 10^{20}$ \cm): 2 coinciding with strong
      metal systems in the nearby line of sight and 1
      matching the emission redshift 
      of the paired QSO; plus 7 Lyman limit systems: 4
      forming two matching couples and 3 without a
      corresponding metal system within $\sim 3000$
      \kms\ in the coupled line of sight.
      In summary, we detect five out of ten
      matching systems within 1000 \kms, indicating a
      highly significant overdensity of strong absorption
      systems over separation lengths from $\sim 1$ to 8
      $h^{-1}$ Mpc.  
      The observed coincidences could arise in 
      gas due to starburst-driven superwinds associated
      with a quasar or a galaxy, or gas belonging to
      large scale structures like filaments or sheets. We
      also 
      determine chemical abundance ratios for three
      damped \Lya\ systems. In particular, for the damped
      system 
      at $z\sim 2.53788$ in the spectrum of Q2344+1228,
      new estimates of the ratios O/Fe, C/Fe are
      obtained: [C/Fe]~$<0.06$, [O/Fe]~$<0.2$. They
      indicate that O and C are 
      not over-solar in this system.   
}

   \maketitle

%

\section{Introduction}

Cosmological simulations based on CDM models 
predict that the {\em forest} of \huno\ \Lya\
absorption lines, observed in QSO spectra, originates in
the fluctuations of the underdense and moderately
overdense regions of the intergalactic medium 
\citep[e.g.][]{cen94,ppj95,zhang95,hernquist96,miralda96,theuns98}. 
The high \huno\ column density systems (Lyman limit and
damped \Lya\ systems), on the other hand, arise from radiatively
cooled gas in galaxy-sized halos \citep[e.g.][]{katz96}. 

In the past few years, the association of high column
density absorption systems ($N(\mbox{\huno}) \gsim
10^{16}$ \cm) with galactic objects has been widely
verified at redshifts up to $z \sim 1$, by direct imaging 
of QSO fields and follow-up spectroscopy. 
The observed impact parameters for galaxies giving rise 
to \mgd\ absorption systems suggest the
presence of extended gaseous halos of spherical geometry
and radii $R \sim 50\ h^{-1}$ kpc (where  $h$ is the
Hubble constant in units of 75 \kms\ Mpc$^{-1}$, and
$q_0=0$) \citep{berg91,berg92,steidel94,guill}. 
While damped \Lya\ systems (DLASs) are likely due to
smaller structures 
\citep{wolfe92,lebrun97}.  


The correlation properties of absorbers
along the line of sight (LOS) were studied
recently. 
A trend of increasing correlation signal with increasing
\huno\ column density at $z \sim 2$ is detected for
QSO absorption lines up to $N($\huno$) \sim 10^{17}$ \cm\
\citep{cristiani97}.  
At the same redshift, higher column density systems are
expected to be more correlated according to the
hierarchical clustering scenario, as they are believed to
be associated with galactic or proto-galactic structures. 
The classic approach to compute the correlation function
is complicated by their rareness.  
In the hypothesis that DLASs are indeed
galaxies, \citet{wolfe93} handles this problem  by
comparing the density of \Lya\ emitters in the field and 
at the redshift of observed DLASs ($<z> = 2.6$), with
that of randomly chosen fields at similar redshift. 
A Poissonian distribution of galaxies in the fields
centred on DLASs is ruled out with more than 99.5 \% 
confidence, but little else can be said
on the correlation function. 

Close pairs or groups of QSO LOSs represent an
alternative, efficient tool to investigate the
correlation properties of absorbers.   
\citet{fran:hew93} find two candidate DLASs in the
spectrum  of Q2138-4427 at $z_{\rm a} \simeq
2.38$ and $2.85$ matching in redshift two weaker \Lya\ 
absorptions in the spectrum of the companion quasar 
Q2139-4434, at a separation of 8 arcmin on the plane of
the sky. Later deep imaging of the field of Q2139-4434
has indeed confirmed the presence 
of a group of red, radio quiet galaxies at $z
\simeq 2.38$. 
This galaxy cluster, with mass $\gg 3 \times 10^{11}\ 
M_{\odot}$, could have collapsed before redshift 5
\citep{francis96,fwd97,francis01a}.  

In this paper, we use two QSO pairs and a triplet to
analyse the correlation behaviour of high matter density
peaks.  We assume that {\em high matter density peaks}
are traced by optically thick absorbers
(i.e. with column density $N($\huno$) \gsim 2\times
10^{17}$ \cm) and by  strong
metal systems (characterised by \cq\ rest equivalent
width $W_{\rm r}(\lambda1548) \ge 0.5$ \AA). 
    
The structure of the paper is the following: Sect.~2
describes the observations and data reduction of 6 new
UVES spectra of three QSO pairs (Q2344+1228 and
Q2343+1232, UM680 and UM681, Q2139-4433 and Q2139-4434); 
in Sect.~3, we describe in more detail one sub-damped
and two damped \Lya\ systems detected in the spectra,
with a particular attention to chemical abundances. 
Section~4 is dedicated to the description of the observed
coincidences.  
The discussion is reported in Sect.~5 and the summary of
results in Sect.~6.
  
All through the paper, we adopt a cosmology with $q_0 =
0.5$ and $h = H_0 / 75$ km s$^{-1}$ Mpc$^{-1}$. 
Spatial separations are always {\em comoving}. 

\section{Observations and data reduction}

In September 2000, we obtained high resolution spectra of
three QSO pairs with the UV and Visual Echelle Spectrograph
\citep[UVES,][]{dekker00} mounted on the Kueyen telescope of the
ESO VLT (Cerro Paranal, Chile). 
The journal of observations is reported in
Table~\ref{obs}. 

\begin{table}
\begin{center}
\caption{Journal of observations September
2000}\label{obs} 
\begin{tabular}{lllll}
\hline
&&&& \\
Object & Mag & $z_{\rm e}$ & Wvl. range & t$_{\rm exp}$ \\
& & & (nm) & (sec) \\
\hline
&&&& \\
UM680 & 18.6 & 2.144 & 305-387 / 477-680 & 7950 \\
UM681 & 19.1 & 2.122 & 305-387 / 477-680   & 11600 \\
Q2344+1228 & 17.5 & 2.773 & 376-498 / 670-10$^3$ & 3600 \\
Q2343+1232 & 17.00 & 2.549 & 376-498 / 670-10$^3$ & 3600 \\
Q2139-4433 & 20.18 & 3.220 & 413-530 / 559-939 & 9000 \\
Q2139-4434 & 17.72 & 3.23 & 413-530 / 559-939 & 7200 \\
\hline
\end{tabular}
\end{center}
\end{table}

Spectra were taken in dichroic mode with a slit of
1.2'' and binning of 2x2 pixels. A binning of 3x2 pixels
was adopted for one of the spectra of the faintest object
Q2139-4433. The resolution is 
$\sim 37000$ and $\sim 35000$ in the blue and in the red
portion of the spectra respectively.  
Wavelength ranges in the blue arm were chosen in order to 
cover most of the \Lya\ forest of each object. Another
paper will be devoted to the detailed discussion of the
lines in this region (D'Odorico et al. in preparation).

Data reduction was carried on by using the specific
UVES pipeline \citep[see][]{balle00} in the framework of  
the 99NOV version of the ESO reduction package, MIDAS.  
The continuum was determined by manually selecting
regions not affected by evident absorption and by
interpolating them with a spline function of 4th degree.   

Metal absorption systems were detected, in general, by
first identifying \cq\ or \mgd\ doublets and then looking
for other ionic transitions at the same redshift. Atomic
parameters for the lines were taken from
\citet{verner}. New oscillator strengths were adopted for
most of the \fed\ transitions \citep{berg94,berg96,raas98}. 
Lines were fitted with Voigt profiles in the LYMAN
context of the MIDAS reduction package.  
The reported errors on column densities are the
$1\,\sigma$ errors of the fit computed in MIDAS. They
possibly underestimate the real error on the
measure since they do not take into account the
uncertainty on the continuum level determination.  
Furthermore, they are the result of a single fitting
model which is not univocal in certain column density
regimes and for heavily blended systems
\citep{font:ball}.  
In the cases in which the column density was weakly 
constrained and the MIDAS procedure could not converge to
a unique solution (e.g. for saturated \huno\ \Lya\ lines),
indicative values of the column density were obtained by
use of the interactive fitting program XVoigt \citep{xvoigt}. 
  
We could analyse  also the UVES spectrum of Q2138-4427
($B = 18.9$, $z_{\rm e} \simeq 3.17$), with similar
resolution and wavelength coverage.   
The detailed description of its reduction and of the two
DLASs present in it, will be given elsewhere (Ledoux et
al. in preparation).


\section{Properties of the main absorption systems}

In the following, we briefly discuss 
the relative abundances of some chemical elements for two
DLASs and a sub-DLAS detected in the present spectra. 
All the abundances are given relative to the solar values
of \citet{grev:and} and \citet{grev:noel}, in the
notation [X/Y]~=~log~(X/Y)$_{\rm
obs}$~-~log~(X/Y)$_{\odot}$ (see Table~\ref{solar}).   
A summary of the obtained chemical abundances is reported
in Table~\ref{abund}. 

\begin{table}
\begin{center}
\caption{Adopted solar abundances for the relevant
chemical elements}\label{solar}
\begin{tabular}{llr}
\hline
&& \\
H & Hydrogen & 0.00 \\
C & Carbon & -3.45 \\
N & Nitrogen & -4.03 \\
O & Oxygen & -3.13 \\
Si & Silicon & -4.45 \\
S & Sulphur & -4.79 \\
Fe & Iron & -4.49 \\
&& \\
\hline
\end{tabular}
\end{center}
\end{table}

\begin{table*}
\begin{center}
\caption{Measured relative chemical abundances. If not
otherwise stated the error on the logarithmic column
densities and on the relative abundances is
0.1}\label{abund}  
\scriptsize{
\begin{tabular}{ccrccccccccc}
\hline
&&&&&&&&&&& \\
Object & Redshift & $\Delta v^{\rm a}$ & log
$N$(\huno)$^{\rm b}$ & [Fe/H] & [Si/H] & [N/H] & [S/H] & 
[N/S] & [C/Fe] & [0/Fe] \\
&&&&&&&&& \\
\hline 
&&&&&&&&& \\
UM681 & 1.78745 & $-$150.5  & 18.6 & & & $-0.6\pm0.2$
& $-0.3\pm0.2$ & &&\\
\hline \\
Q2343+1232 & average$^{\rm c}$ & & 20.35  &$-1.2\pm 0.2$ &
& $-$1.1 & $-$0.7 &&& \\     
	& 2.43125 & 0.0 & & & & && $-$0.3 &&\\     
\hline \\
Q2344+1228 & average$^{\rm c}$ & & 20.4 & $-1.8\pm 0.2$
& $-$1.85 & $-$2.75 &&&& \\ 
         & 2.53746 & $-$35.6 & && & & & &  $<$0.06 & $<$ 0.2 \\
\hline   
\end{tabular}
}
\end{center}
\scriptsize{$^{\rm a}$ Relative velocities as reported in the
corresponding figures \\
$^{\rm b}$ Since it is not possibile to disentangle the
velocity structure of the \huno\ \Lya\ absorption lines,
the profiles have been fitted with one or two components
at the redshift of the stronger components observed in
the neutral and singly ionised absorption lines \\  
$^{\rm c}$Average value obtained considering the sum of
the column densities of all the components of the iron
absorption profile }
\end{table*}

\subsection{The sub-DLAS at $z_{\rm a}\sim$~1.788 in UM681} 
\begin{figure}
   \begin{center}
	 \resizebox{\hsize}{8cm}{\includegraphics{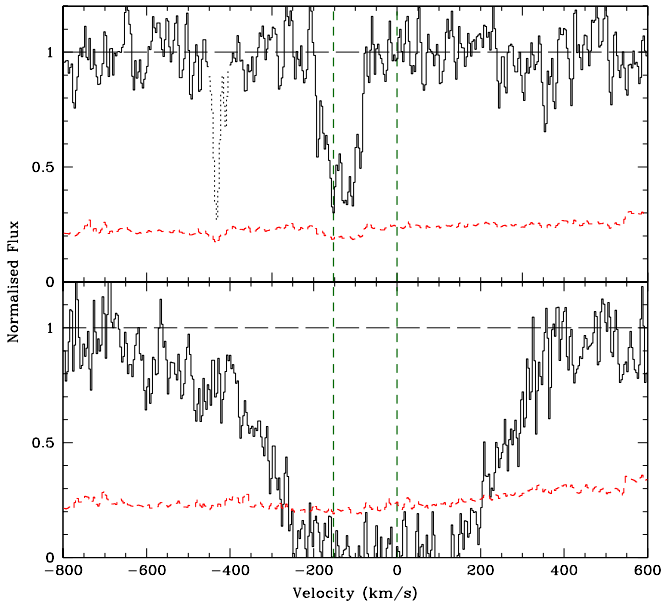}}
   \end{center}
   \caption{$z \sim 1.788$ - Coincidence between the
	 sub-damped \Lya\ absorption line at $z_{\rm
	 a}\simeq 1.788$ in the spectrum of UM681 (bottom
	 panel) and a weak \Lya\ absorption line ($\log
	 N($\huno$) \simeq 13.81 \pm 0.05$) in the
	 spectrum of UM680 (top panel). 
	 The spatial separation between the two
	 LOSs at this redshift is $\sim 870\ h^{-1}$ kpc.   
         The dotted lines mark the position of the 2
	 components fitting the sub-DLAS at redshifts
	 $z_{\rm a}=1.78745$ and 1.78885 (origin of the
	 velocity axes)}  
\label{hit178}
\end{figure}
\begin{figure}
   \begin{center}
	 \resizebox{\hsize}{11cm}{\includegraphics{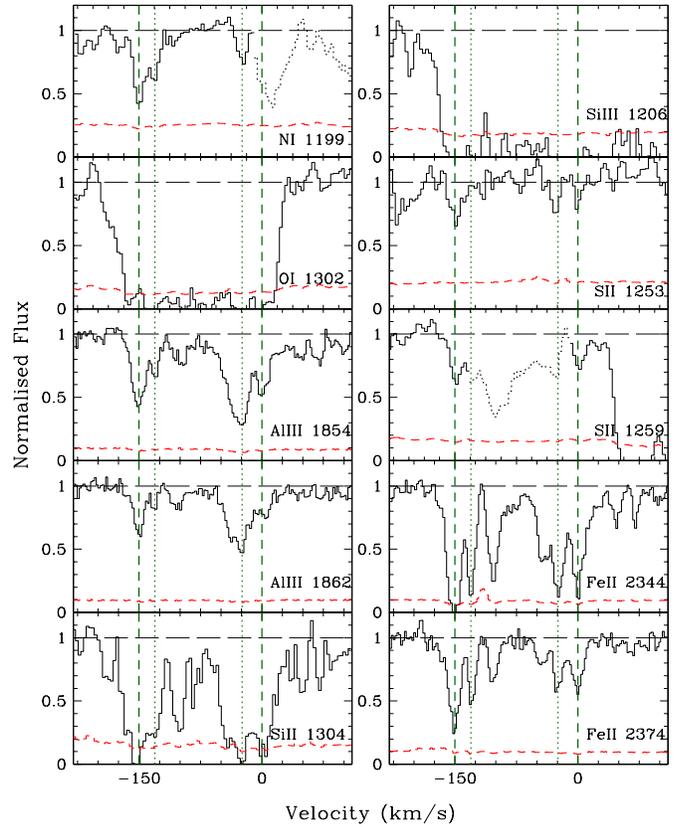}}
   \end{center}
   \caption{UM681: ionic transitions of the system
	 at $z_{\rm a} \simeq 1.788$. The dotted vertical
	 lines mark the position of the components
	 discussed in the text. The
	 two thick, dashed lines corresponds to the
	 redshifts of the two 
	 components used to fit the \huno\ \Lya\ absorption} 
\label{sys178}
\end{figure}
This is a newly detected metal absorption system.  
The \huno\ \Lya\ absorption at this redshift shows a 
clear Lorentzian wing on the blue side of the velocity
profile (see Fig.~1). 
In order to obtain an estimate of the \huno\ column
density, we fit the profile with two components at
the redshifts of the two strongest groups of lines seen in
the singly ionised element profiles (see Fig.~\ref{sys178}). 
At $z_{\rm a} \simeq 1.78745$ ($v \simeq -150$ \kms\ in
the figures), we measure a column density $\log 
N$(\huno)$\sim 18.6\pm0.1$ and at $z_{\rm a} \simeq
1.78885$ ($v \simeq 0$ \kms), $\log N$(\huno)$\sim 19.0
\pm 0.1$.  
The associated heavy element transitions have a complex
structure spread over about 250 \kms\ with at
least 8 narrow components in \fed, \sidue\ and \altre\
and 4 components in \nuno\ and \sdue\ 
(see Fig.~\ref{sys178}). \ouno, \cdue\ and \sitre\ are
either saturated or blended, \siq, \cq\ and \aldue\ are
outside our wavelength range. 

The observed \huno\ column density is not high enough to
assure that no ionisation corrections have to be applied
to get the relative chemical abundances \citep[see
e.g. ][]{viegas}.   
We derive an indicative measure
of the nitrogen and sulphur abundances by using the
Cloudy software package \citep{cloudy} to 
build a photoionisation model and estimate the ionisation
corrections. We consider the UV background flux due to
quasars and galaxies \citep{MHR99} and try to recover the
observed column densities of the components at $z_{\rm a} 
\simeq 1.78745$ and 1.78765 ($v \simeq -150$ and
$-$130~\kms\ in Fig.\ref{sys178}). The results do not
change significantly if we adopt an ionizing spectrum due
to a single stellar population of solar metallicity and
0.1 Gyr. 
The resulting ionisation corrections are $\sim$~2.3, 0.2
and 14 to be multiplied for the ratios  \nuno/\huno,
\sdue/\huno\ and \nuno/\sdue\ respectively.   
Since the corrections are relatively small in the first two
cases, we compute the corresponding abundance ratios. 
\nuno\ $\lambda\,1199$ is partially
blended (see Fig.~\ref{sys178}), thus we estimate the
nitrogen abundance from the two components at lower redshift
($v \simeq -150$ and $-$130~\kms\ in Fig.\ref{sys178}),
associated with the $z_{\rm a} \simeq 1.78745$ \huno\
component with $\log N($\huno$) \simeq
18.6\pm 0.1$. From this we derive [N/H]~$\simeq -0.6 \pm
0.2$ corrected for the ionisation. This value is about
one order of magnitude larger than the higher value
measured in DLASs published in the literature
\citep{miriam98,lu98}.      
The corrected sulphur abundance ratio for the same
components is [S/H]~$\sim -0.3 \pm 0.2$ which again is
about one order of magnitude larger than what is observed
in DLASs. 

These measurements, although slightly uncertain, suggest
that sub-DLAS could have higher metal abundances than
DLASs, as already observed in LLS
\citep[e.g.][]{dodo01}, and probe a more evolved
chemical stage of high redshift galaxies when gas has
been partly consumed by star formation.

\begin{figure}
   \begin{center}
	 \resizebox{\hsize}{11cm}{\includegraphics{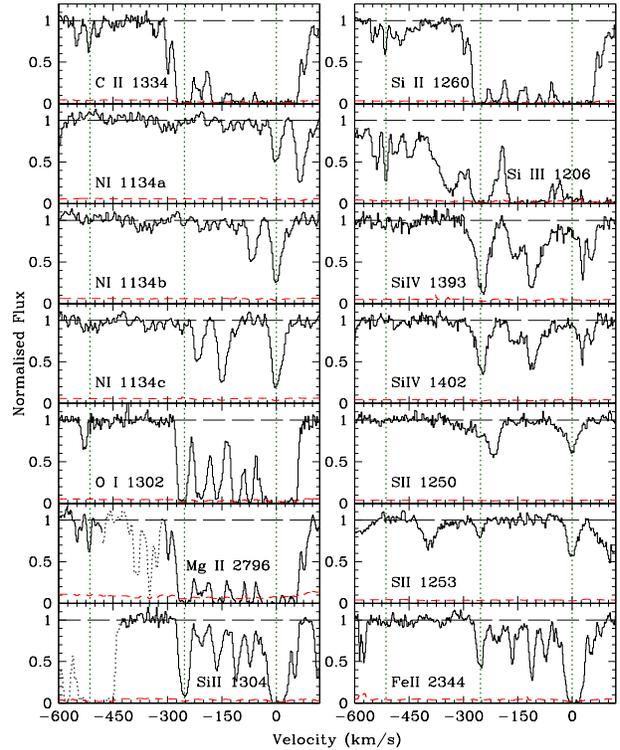}}
   \end{center}
   \caption{Q2343+1232: ionic transitions
	 associated to the DLAS at $z_{\rm a} \simeq 
	 2.43125$. The dotted line at the extreme left
	 marks the position of the satellite sub-system at
	 $z_{\rm a} = 2.42536$ (see text). The dotted lines
	 on the right are drawn at the redshifts $z_{\rm
	 a} =2.42834$ and 2.43125
	 (origin of the velocity axes) of the  
	 components discussed in the text}  
\label{dla243}
\end{figure}

\subsection{The DLAS at $z_{\rm a}\simeq$~2.43125 in Q2343+1232}

This DLAS, as well as the one seen along the LOS of
Q2344+1228 (see next section), has first been detected by 
\citet{sarg87}. 
The two QSOs were observed recently with HIRES+Keck and
the relative chemical abundances of the two systems were
used in statistical samples
\citep{rauch97,lu98,pw98,pw01,prochaska01}.   
    
The metal absorption complex corresponding to this damped 
system  is made of two groups of lines. 
The major one counts at least 8 components, with the
strongest one at $z_{\rm a} \simeq 2.43125$ ($v =
0$~\kms\ in Fig.~\ref{dla243}).  
This component is heavily saturated
in \cdue, \ouno, \mgd, and \sidue. It shows absorption
due to the two triplets of \nuno, $\lambda\,1134$ \AA\
and  $\lambda\,1200$ \AA, and to the \sdue\ triplet,
$\lambda\lambda\lambda\,1250,1253,1259$ \AA. 
The \siq\ doublet is clearly identified in this complex,
the \cq\ doublet is outside our spectral range but was
detected by \citet{sbs88}.  
A satellite sub-system is observed at more than 500~\kms\ 
from the centre of the main one, at $z_{\rm a} \simeq
2.42536$.  It is very weak and shows
transitions due to \cdue, \mgd, \sidue\ and \sitre. 

Since we cannot disentangle the velocity structure of the 
hydrogen absorption, we assume a single component at the 
redshift of the strongest component observed in singly
ionised lines at $z_{\rm a} \simeq 2.43125$. 
The total \huno\ column density is $\log N($\huno$)
\simeq 20.35 \pm 0.05$ and the error is mainly due to the
uncertainty in the position of the continuum.
The average iron abundance is [Fe/H]$\simeq -1.2\pm 0.2$.
While, we obtain [S/H]~$\simeq -0.7 \pm 0.1$ and
[N/H]~$\simeq -1.1 \pm 0.1$.    
Those estimates are $\sim 0.2$ and 0.5 dex higher
respectively, than those reported by \citet{lu98}. 
The difference for sulphur is within the uncertainties (at
the $3\,\sigma$ level), while the larger one for nitrogen
could be due to the fact that \citet{lu98} used the
saturated \nuno\ triplet at $\lambda\,1200$ \AA. 

As previously stated, the main component is badly
saturated for all the observed transitions due to C, O
and Si, so relative abundances for these elements cannot
be determined. On the other hand, we can study the
abundance ratios of S, N and Fe at this redshift,
assuming that ionisation corrections are negligible. This
hypothesis is supported by the large column density
characterising this component and by the absence of \siq\
absorption. 
We measure column densities:
$\log N($\sdue$)\simeq 14.75\pm0.05$, $\log
N($\nuno$)\simeq 15.16\pm0.05$ (where we do not 
consider the \nuno\ triplet at $\lambda\,1200$ \AA\
because it is saturated and affected by the wing of the
damped \huno\ \Lya\ line) and $\log N($\fed$)\simeq
14.49\pm 0.08$. From which we derive the abundance
ratios: [S/Fe]~$\simeq 0.6\pm 0.1$, and [N/S]~$\simeq
-0.3\pm 0.1$. The latter abundance ratio, which is not
affected by dust, can be compared with the ratios [N/O]
measured for metal poor Galactic stars making the
assumption, [S/O]~$\equiv 0$ as reported by
\citet{miriam98}.     
Our measurement is larger than any other for DLASs
present in the literature \citep[see also][]{lu98} and it
is consistent with values obtained for \ion{H}{ii}
regions in dwarf galaxies. 

The strong \siq\ absorption imply that ionisation
corrections could be necessary to determine the abundance 
ratios in the other components of the system.  
Nothing can be said on the \huno\ column density
corresponding to the single components. This makes hard
the realization of a photoionisation model to determine
the ionisation corrections.  
We report the measured column densities of sulphur and
nitrogen (a faint absorption is observed for the
transition \nuno\ $\lambda\,1200$) relative to the
component at $z_{\rm a} \simeq 2.42834$ ($v \simeq -250$
\kms\ in Fig.~\ref{dla243}), $\log N($\sdue$)\simeq
14.22\pm0.05$ and $\log N($\nuno$)\simeq 13.5\pm0.05$. 



\begin{figure}
   \begin{center}
	 \resizebox{\hsize}{11cm}{\includegraphics{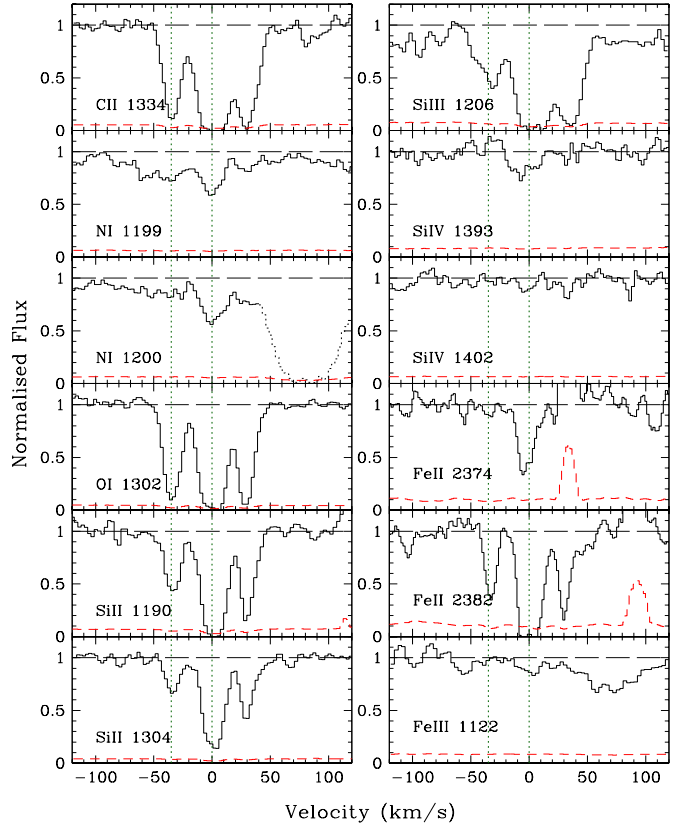}}
   \end{center}
   \caption{Q2344+1228: ionic transitions associated to
	 the DLAS at $z_{\rm a} = 2.53788$ (origin of the
	 velocity axes). The left dotted line marks the
	 position of the component at $z_{\rm a} =
	 2.53746$ discussed in the text}  
\label{dla253}
\end{figure}

\subsection{The DLAS at $z_{\rm a}\simeq$~2.53788 in
	 Q2344+1228} 

The damped \huno\ \Lya\ absorption line of this system
has been fitted with a single component at the redshift
of the strongest component observed in the neutral and
singly ionised lines of associated heavy elements
($z_{\rm a} \simeq 2.53788$). 
The \huno\ column density is $\log N($\huno$) \simeq 20.4
\pm 0.1$, where the error on the column density is due
mainly to the positioning of the continuum level. 

In Fig.~\ref{dla253}, we show the ionic transitions
observed for the DLAS. The \cq\ doublet is outside our
wavelength range and it was not observed in the low
resolution spectrum by \citet{sbs88}. 
The iron and silicon column densities can be derived from
non saturated lines to obtain the average values,
[Fe/H]~$\simeq -1.8 \pm 0.2$, and [Si/H]~$\gsim -1.85 \pm  
0.1$.  While, [N/H]~$\simeq -2.75 \pm 0.11$, which is in
good agreement with the value found by \citet{lu98}. 

The absence of high ionisation lines and the simple
velocity profile of the system allow the assumption that
ionisation corrections are negligible in this case. We
can thus obtain reliable abundance ratios from the column
densities of the transitions observed in the single
components. 
 
\nuno\ is observed in the central component, we compute
the abundance ratio [N/Fe]~$\simeq -0.8 \pm 0.1$, which
is consistent with previous measurements for DLAS. 
The central component is unusable to derive reliable
abundance measures for other chemical elements because
all the observed lines are heavily saturated. 

In the component at lower redshift ($z_{\rm a} \simeq
2.53746$, $v\simeq -35.6$ \kms\ in Fig.~\ref{dla253}), 
the ratio [\sitre/\sidue]~$\simeq -0.66\pm 0.07$ 
implies that ionisation corrections are small at this
velocity. We observe transitions due to \sidue\ and \fed\ 
that are not saturated, from which 
the relative abundance [Si/Fe]~$\simeq 0.2 \pm 0.1$ is
obtained. On the other hand, \cdue\ and \ouno\ are
slightly saturated but not going to zero, this results in
the upper limits: [C/Fe]~$\lsim 0.06 \pm 0.07$ and  
[O/Fe]~$\lsim 0.2 \pm 0.1$.   
Reliable measures of O and C abundances are quite
rare. We discuss the implications of our result in the
following section. 

\subsection{Comments on the measures of [O/Fe] and [C/Fe]}

$\alpha$-capture elements are mainly produced by Type II
SNe which should dominate in the early stages of the
chemical evolution of galaxies, while Type I SNe
contribute iron peak elements later on. Therefore, the
[$\alpha$/Fe] abundance ratio can be used to trace the
chemical evolution history and, to a certain extent, the
nature of galaxies.   
Oxygen and sulphur are more reliable estimators of
$\alpha$-element abundances than silicon which is subject
to dust depletion. 
Abundance studies of carbon \citep[e.g.][]{tomkin95}
indicate that in the disk of our Galaxy [C/Fe] and
[$\alpha$/Fe] show similar trends with [Fe/H].

Measures of C and O in DLASs are generally complicated by
the fact that often the only available lines are
\cdue\ $\lambda\,1334$ and \ouno\ $\lambda\,1302$ which
most of the times are heavily saturated. 
In the DLAS described in Sect.~3.3, we constrain
the values of the O/Fe and C/Fe ratios, considering
a single component which is only mildly saturated and not
going to zero. 
We derive that the C/Fe ratio is consistent with
solar while the O/Fe and Si/Fe ratios are consistent
among them and show a very small enhancement. 
The average iron abundance of the system is about 1/100
solar.     
Our result, together with the recent measures by
\citet{molaro00} for the DLAS in the spectrum of Q0000-26, 
indicates that there is no evidence for the [O/Fe] ratio
to be over-solar in DLAS. This is at a variance with
what is observed in the atmosphere of Galactic stars at
the same metallicity \citep[but see also][]{mirka01}.

The abundance pattern which is closest to the above data
is that of an old starburst, as is observed at the
boundaries of our galactic disk, although, in general, for 
larger iron abundances  \citep{chiappini99}.

\section{Coincidences of high matter density peaks}


In this section, we describe the observed pairs of
quasars and list the absorption systems with
$N($\huno$) \gsim 2\times 10^{17}$ \cm\ found in the 7
spectra. For each of the systems, we search the adjacent
line of sight for the presence of any absorption at the
same redshift. When a LLS or a \cq\ system with rest 
equivalent width $W_{\rm r} > 0.5$ \AA\ is seen along the
second LOS within $\sim 1000$ \kms\ from the former LLS,
we call this a coincidence. 
The observed number densities of LLS and \cq\ absorption
systems with $W_{\rm r} > 0.5$ \AA\ are similar at
the same redshift \citep[e.g.][]{ssb88,steidel90}. 
We can therefore assume that they trace the same kind of
overdensity. 

The numbers associated with the coincidences correspond
to those in Table~\ref{tab:hits} and  Fig.~\ref{hits}. 


\subsection{The QSO pair UM680 and UM681}

These two QSOs (also called Q0307-195A,B) are separated
by 56 arcseconds on the plane of the sky, corresponding
to $\sim 830-940\ h^{-1}$ kpc in the considered redshift
interval.  
Spectra at low and intermediate resolution of this pair
have been used in the past to study the correlation of
\cq\ and \Lya\ forest lines \citep{sha:rob83,dodo98}. 

\vskip 12pt

\noindent
1) $z \simeq 1.788$ -   
There is no metal system along the LOS of UM680
corresponding to the sub-DLAS at $z_{\rm a} \simeq 1.788$
observed in the spectrum of UM681 (see Sect.~3.1). A weak
\Lya\ absorption, $\log N($\huno$) \simeq 13.81 \pm
0.05$, is observed at $z_{\rm a} \simeq 1.7876$ (see 
Fig.~\ref{hit178}). From the observed number
density of \huno\ \Lya\ absorption lines with column
density in the range $13.1 \le \log N($\huno$) \le 14$
\citep{kim01}, the probability for such an absorption to
fall in a velocity bin of 200 \kms\ at this redshift is
${\cal P} \sim 0.3-0.4$.

\begin{figure}
   \begin{center}
	 \resizebox{\hsize}{8cm}{\includegraphics{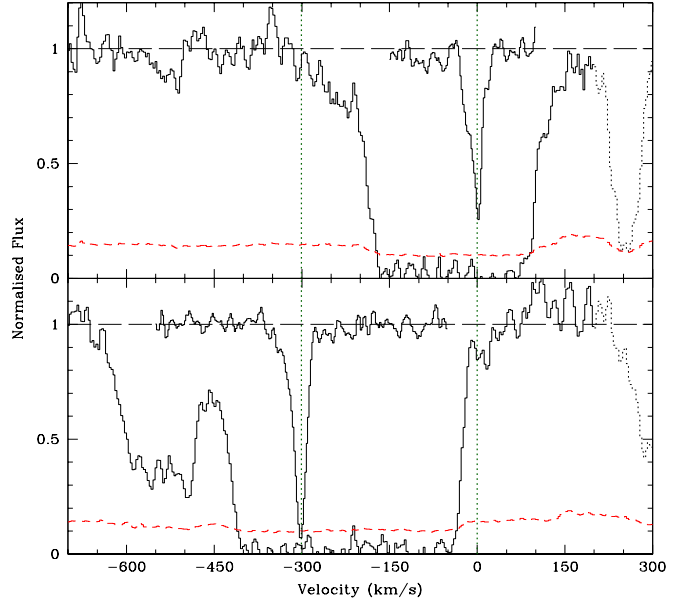}}
   \end{center}
   \caption{$z \sim 2.03$ - Coincident absorption systems in the spectra
	 of UM680 (top panel) and UM681 
	 (bottom panel). The \huno\ \Lya\ transitions
	 are shown with superposed the corresponding 
	 \aldue\ $\lambda\,1670$ one. The transverse
	 spatial separation between the two LOSs at this
	 redshift is $\sim 924\ h^{-1}$ kpc. The dotted
	 lines mark the position of the metal lines at
	 $z_{\rm a} = 2.03215$ and $z_{\rm a} = 2.03520$
	 (origin of the velocity axes)}    
\label{hit203}
\end{figure}

\vskip 12pt
\noindent
2) $z \simeq 2.03$ - 
The coincident systems observed at $z_{\rm a} \simeq
2.03520$ and  $z_{\rm a} \simeq 2.03215$ in the spectra
of UM680 and UM681 respectively, are two candidate LLSs
which show absorption lines due to the same ionic
transitions with a shift of $\simeq 300$ \kms\ (see
Fig.~\ref{hit203}).  
We detect low ionisation absorption lines due to
\aldue, \sidue\ and \fed, together with \altre, \sitre\
and \fetre, the latter only in UM680. The corresponding
\siq\ and \cq\ absorption doublets are outside our
wavelength coverage, but their presence is discussed in
\citet{sha:rob83}.  

It is not possible to constrain the value of the \huno\
column density of both systems due to the complexity of
the profile. The \Lyb\ lines are in a region of the
spectrum with low signal-to-noise ratio and
probably blended.  
From the equivalent width ratio of \sidue\ and \fed\
to \cq\ \citep[as measured by][]{sha:rob83} we derive 
that the systems are likely in a low excitation state and
have $N($\huno$) > 10^{18}$ \cm\ \citep[see][]{bs86}.

\begin{figure*}
   \begin{center}
	 \resizebox{\hsize}{10cm}{\includegraphics{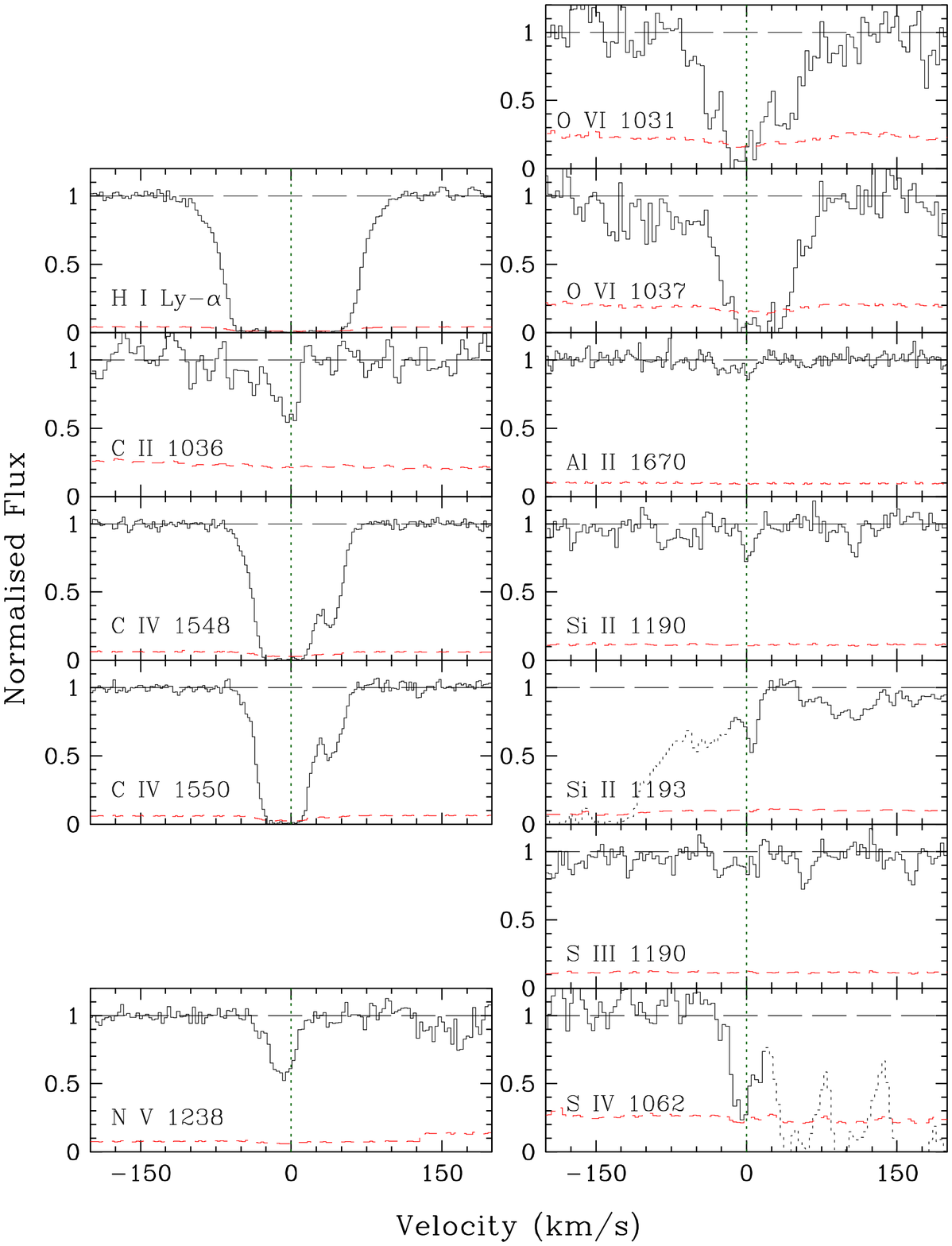}
          \includegraphics{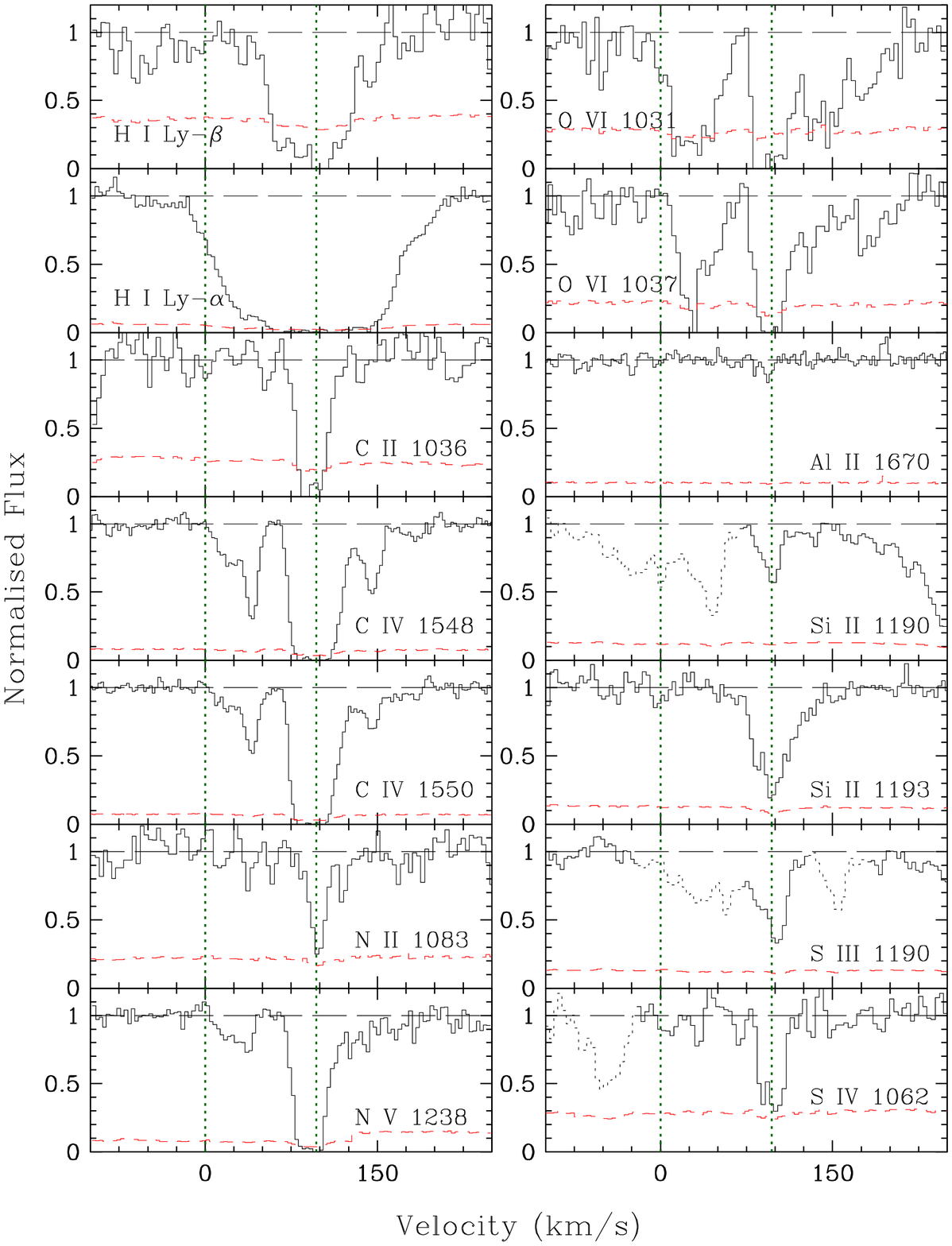}}
   \end{center}
   \caption{Left: Ionic transitions observed at  $z_{\rm a} =
	 2.12209$ (origin of the velocity axes) in the
	 spectrum of UM681. This 
	 redshift is obtained from the fitting of the low
	 ionisation lines; the center of the velocity
	 profile of the high ionisation lines, determined
	 by the \nc\ $\lambda\,1238$ \AA\ transition, is
	 shifted by $\simeq -8.5$ \kms. Right: 
	 Ionic transitions observed at $z_{\rm a} = 
	 2.12312$ ($v \simeq 97$ \kms) in the spectrum of
	 UM680. The origin of the velocity axes is kept 
	 at $z_{\rm a} = 2.12209$, redshift of the
	 low ionisation transitions observed in the
	 coincident system in the spectrum of UM681. The
	 transverse separation between the two LOSs at
	 this redshift is $\sim 940\ h^{-1}$ kpc}   
\label{sys212}
\end{figure*}

\vskip 12pt
\noindent
3) $z \simeq 2.122$ - 
The QSO UM681 presents a metal system at its emission
redshift ($z_{\rm a} \simeq 2.12209$) with lines due to 
\cq, \nc, \osei\ and \ion{S}{iv} and also weak
low ionisation lines (see Fig.~\ref{sys212}). 
This system, although characterized by highly ionised
transitions, has a velocity spread of less than
$\sim 250$ \kms\ and does not show any evidence of
partial coverage. 
Furthermore, the presence of singly ionised absorption
lines and the symmetric velocity profile favour an
absorber with a dense core. 
Therefore, although the system is located in the vicinity
of the quasar it is probably not associated with it.

In addition, there is a very similar absorption
system along the LOS of UM680, at
$z_{\rm a} \simeq 2.12312$, corresponding to a velocity  
shift of  $\sim 100$ \kms (see Fig.~\ref{sys212}). 
The transverse spatial separation between the two LOSs at
this redshift is $\sim 940\ h^{-1}$ kpc.   
The latter system is located at $\sim 2000$ \kms\ from the
emission redshift of UM680; the same arguments as before
are valid to reject the hypothesis that this is due to gas
associated with either of the two quasars. 
The observed \huno\ \Lya\ and \Lyb\ absorption lines 
for this system are consistently fitted with a main
component of column density $\log N($\huno$) \gsim 
17.3$.   

\begin{figure}
   \begin{center}
	 \resizebox{\hsize}{8cm}{\includegraphics{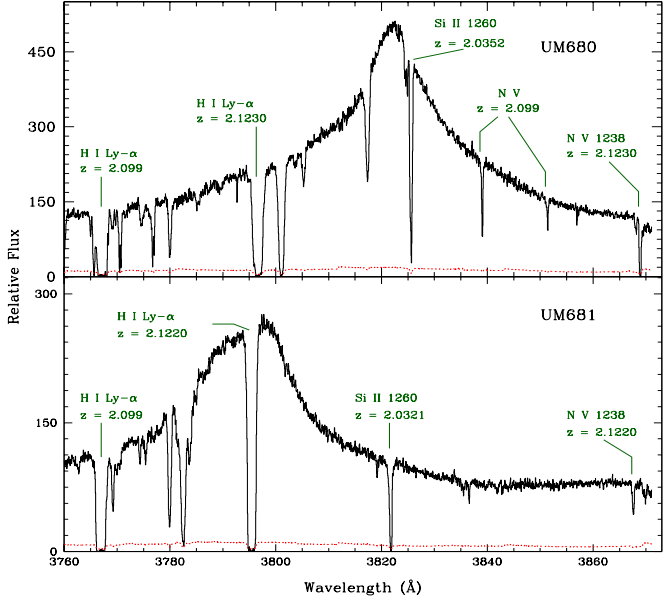}}
   \end{center}
   \caption{\Lya\ emission region in the spectra of the
	 QSOs UM680 (top panel) and UM681 (bottom 
	 panel). Marked are the two coinciding \huno\
	 \Lya\ lines at $z_{\rm a} \simeq 2.099$ and 2.122 and
	 other interesting metal absorption lines (see
	 text). The two LOSs are separated by $\sim 940\
	 h^{-1}$ kpc at $z = 2.122$}   
\label{lya212}
\end{figure}

Figure~\ref{lya212} shows the \huno\ \Lya\ emission
region in the two QSO spectra.  
The coinciding \Lya\ absorptions at $z_{\rm a} \sim
2.122$ are shown, together with the associated \nc\
$\lambda\,1238$ lines (the \nc\ $\lambda\,1242$
transitions fall outside the observed wavelength range).
Another pair of \Lya\ absorptions is observed at $z_{\rm
a} \sim 2.099$, which shows an associated \nc\ doublet in
the spectrum of UM680, while does not have any detected
associated metal line in UM681. 

\citet{sha:rob83} suggest the existence of a uniform,
1 Mpc diameter, gaseous disk associated with UM681 to
explain the coincidence at $z_{\rm a} \sim 2.122$. The
presence of a further coincidence at $\sim 2000$
\kms\ from this one, favours the thesis that
the absorptions are due to a coherent gaseous
structure embedding both quasars and possibly small
galactic objects. 
Deep imaging of the field could possibly shed light on
the nature of the absorbers and of the ionising processes
at work in the gas.


\subsection{The QSO pair Q2344+1228 and Q2343+1232}

The first spectra of this QSO pair were presented by 
\citet{sbs88}, the two objects are separated by 5
arcmin on the plane of the sky, corresponding to a
transverse spatial separation of $\simeq 5\ h^{-1}$ Mpc
in the considered redshift range. 
The remarkable feature is the presence of
a DLAS in each of the LOS (see Sect.~3). 

\begin{figure}
   \begin{center}
	 \resizebox{\hsize}{8cm}{\includegraphics{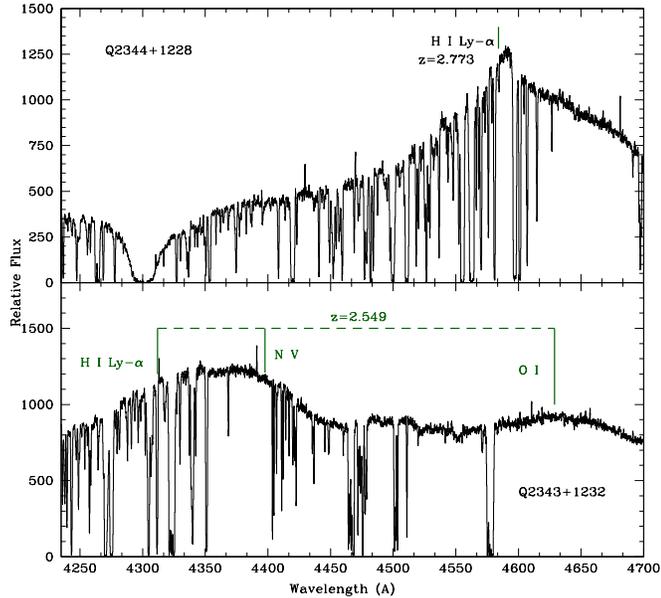}}
   \end{center}
   \caption{\huno\ \Lya\ emission regions in the spectra
	 of Q2344+1228 (top panel) and Q2343+1232
	 (bottom panel). The transverse spatial
	 separation between the two LOSs in this region
	 is $\sim 5\ h^{-1}$ Mpc. 
	 The dashed vertical lines mark: in
	 the top panel, the position of the \huno\ \Lya\
	 emission at $z_{\rm e}=2.773$; in the bottom
	 panel, the labeled emission lines at  $z_{\rm
	 e}=2.549$, with the shifted rest wavelengths by 
	 \citet{tytler:fan} }
\label{lyaem}
\end{figure}

The emission redshift of Q2343+1232 reported by
\citet{lu98}, $z_{\rm e} \simeq 2.549$, is consistent
with the position of the emission lines observed in the
\citet{sbs88} spectrum (\siq+\ion{O}{iv}] and \cq) and
with the \ouno\ emission in our spectrum (marked in
Fig.~\ref{lyaem}), when the shifted rest wavelengths
computed by \citet{tytler:fan} are used. 
Likely, the peak observed at $\lambda \sim 4375$ \AA\ is
partly due to the \nc\ emission, while the maximum of the
\Lya\ emission is strongly absorbed.  
We identify two absorption systems at $z_{\rm a} > z_{\rm
e}$: a \nc\ doublet and the corresponding \Lya\
absorption at $z_{\rm a} \simeq 2.5698$ ($\Delta v \simeq
1750$ \kms), together 
with another possible \Lya\ line at $z_{\rm a} 
\simeq 2.579$ ($\Delta v \simeq 2500$ \kms).  
They do not show any signature of partial coverage and
they could be explained by the presence of a cluster 
of galaxies of which the QSO itself is a member
\citep[e.g.][]{weymann79}.    


\begin{figure}
   \begin{center}
	 \resizebox{\hsize}{8cm}{\includegraphics{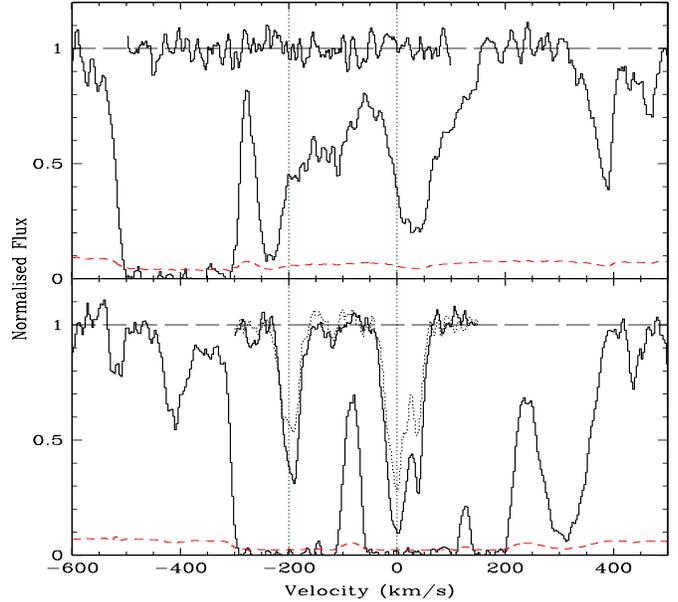}}
   \end{center}
   \caption{$z \sim 2.171$ - Coincidence between the LLS at $z_{\rm a}
	 \simeq 2.17115$ (origin of the velocity axes) in
	 the spectrum of Q2343+1232 
	 (bottom panel) and a \huno\ \Lya\ absorption
	 without associated metals in the spectrum of
	 Q2344+1228 (top panel). Overplotted on the \Lya\
	 absorptions are the corresponding \cq\
	 $\lambda\,1548$ and $\lambda\,1550$ spectral
	 regions showing a detectable absorption only in
	 Q2344+1232. The two LOSs are separated by $5\
	 h^{-1}$ Mpc}    
\label{hit217}
\end{figure}
\vskip 12pt
\noindent
4) $z \simeq 2.171$ - 
In the spectrum of Q2343+1232, we identify a metal system
at $z_{\rm a} \simeq 2.171$ which could be a LLS on the
ground of the column density ratios of the observed
transitions \citep{bs86}. In particular, \ouno/\cq~$\sim
0.6$, \mgd/\cq~$\sim 0.4$ and  \sidue/\siq~$\sim 3.7$. 
No metal system is detected along the companion LOS within
$\sim 3000$ \kms. However, a complex \huno\ \Lya\
absorption is present (see Fig.~\ref{hit217}) whose
velocity profile appears to match that of the \Lya\ 
in Q2343+1232 when shifted red-ward by $\sim 240$ \kms. 

\begin{figure}
   \begin{center}
	 \resizebox{\hsize}{8cm}{\includegraphics{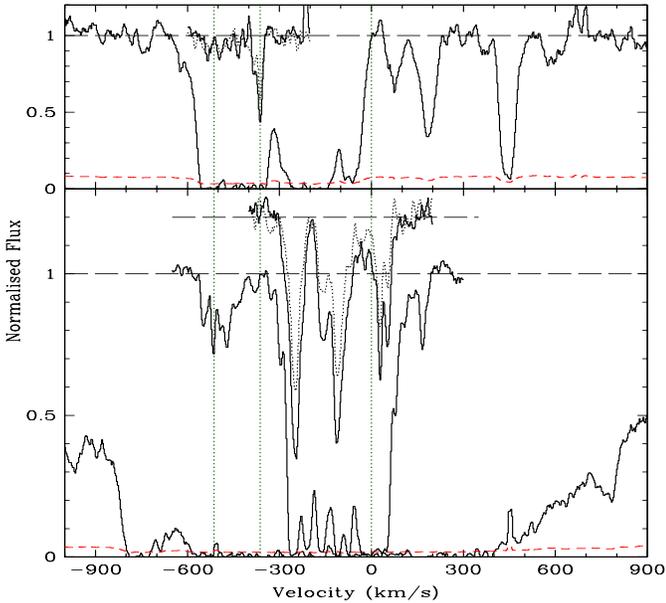}}
   \end{center}
   \caption{$z \sim 2.43$ - Coincident absorption systems
	 in the spectra of Q2344+1228 (top panel) and
	 Q2343+1232 (bottom panel). The two LOS are
	 separated by $\sim  5.3\ h^{-1}$ Mpc. The \huno\
	 \Lya\ transitions 
	 are shown with superposed: the corresponding
	 \siq\ doublet at $z_{\rm a} =2.4271$ (top) and  
	 the \sidue\ $\lambda\,1260$ plus the \siq\
	 doublet shifted upward by 0.2 for clearness 
	 (bottom). The dotted vertical lines mark from
	 left to right: the position of the weak metal
	 complex in Q2343+1232 at $z_{\rm a} =2.42536$,
	 the main component of the metal absorption in
	 Q2344+1228 and the main component of the low
	 ionisation metal lines in Q2343+1232 at $z_{\rm a} =
	 2.43125$ (origin of the velocity axes)}    
\label{hit243}
\end{figure}

\vskip 12pt
\noindent
5) $z \simeq 2.43$ - 
In the spectrum of Q2343+1232, the DLAS at $z_{\rm a}
\simeq 2.43$ (see Sect.~4.2) coincides with a metal
system at $z_{\rm a} \simeq 2.4271$ (redshift of the
\siq\ main component)  in the companion
LOS, showing only high ionisation lines \citep[\cq\ was
detected by ][]{sbs88} 
and a strongly saturated \Lya\ ($W_{\rm r} 
\simeq 2$ \AA) (see Fig.~\ref{hit243}). 
The \huno\ absorption is likely not a LLS
since singly ionised lines are not detected (like \mgd\ 
and \fed). An acceptable fit of the profile is
obtained with two main components at $N($\huno$) \sim 
10^{15}$ and $\sim 6.3\times10^{15}$ \cm, which however
should be considered as lower limits. 

\vskip 12pt
\noindent
6) $z \simeq 2.54$ -    
The DLAS at $z_{\rm a} \simeq 2.53788$ in
the spectrum of Q2344+1228 (see Sect.~4.1) does not have a
corresponding metal system on the LOS to  Q2343+1232, but
it is indeed at $\sim 940$ \kms\ from the \huno\ \Lya\
emission at the redshift of this quasar (see
Fig.~\ref{lyaem}).  

\subsection{The QSO triplet Q2138-4427, Q2139-4433 and
Q2139-4434} 

The quasar Q2139-4434 ($z_{\rm e} = 3.23$) was 
observed at intermediate resolution together with its
companion Q2138-4427 ($z_{\rm e} = 3.17$) by
\citet{fran:hew93}.  
They are separated by 
8 arcmin on the plane of the sky. Francis and  
Hewett observed common strong \Lya\ absorptions at $z
\sim 2.38$ and $z \sim 2.85$ and further imaging of the
field revealed the presence of a cluster of galaxies at
$z \sim 2.38$ \citep{francis96,fwd97,francis01a}.
\citet{wolfetal95} confirmed the damped nature of the
system at $z \sim 2.85$ in the spectrum of Q2138-4427.
We obtained high resolution spectra of  Q2138-4427,
Q2139-4434 and of Q2139-4433 \citep[$z_{\rm e} =
3.220$, $R = 19.97$;][]{hawk:veron96}. The latter two
QSOs are separated by 1 arcmin on the plane of the sky.   

\begin{figure}
   \begin{center}
	 \resizebox{\hsize}{6cm}{\includegraphics{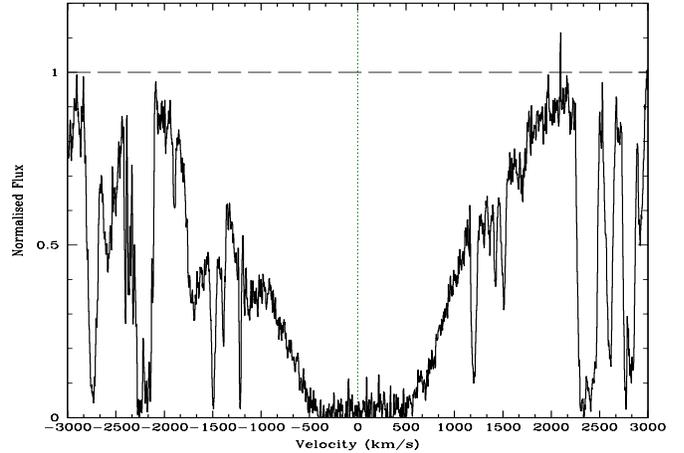}}
   \end{center}
   \caption{Q2138-4427: \huno\ \Lya\ absorption line at
	 $z_{\rm a} = 2.38279$} 
\label{lya238}
\end{figure}

\begin{figure}
   \begin{center}
	 \resizebox{\hsize}{9cm}{\includegraphics{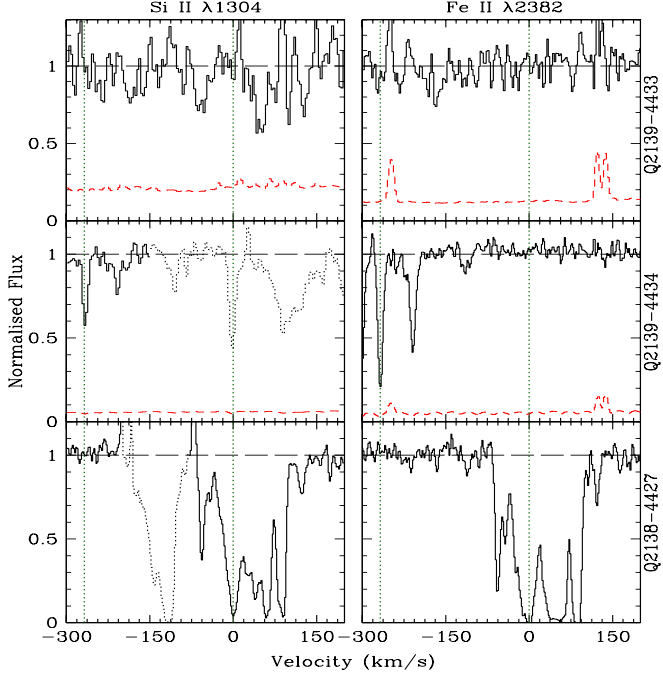}}
   \end{center}
   \caption{$z \sim 2.38$ - Coincident absorption systems
	 in the spectra of Q2139-4434 (mid panels) and
	 Q2138-4427 (bottom panels), with a transverse
	 separation of $\sim 9\ h^{-1}$ Mpc. The left
	 panels show the \sidue\ $\lambda\,1304$
	 transition and the right panels show the 
	 \fed\ $\lambda\,2382$ one. In the top row, the
	 corresponding regions in the spectrum of
	 Q2139-4433 are plotted. At this redshift, the
	 separation between the LOSs to Q2139-4433 and
	 Q2139-4434 is $\sim 1\ h^{-1}$ Mpc and between
	 Q2139-4433 and Q2138-4427 is $\sim 7.7\ h^{-1}$
	 Mpc. The origin of the velocity
	 axes is set at $z_{\rm a} = 2.38279$. The other vertical
	 dotted line marks the position of the system in
	 Q2139-4434 at $z_{\rm a} = 2.37977$}   
\label{hit238}
\end{figure}

\vskip 12pt
\noindent
7) $z \simeq 2.38$ - 
The strong \Lya\ absorption at $z_{\rm a} \simeq 2.38$ in
the spectrum of Q2138-4427 has at least
one visible damped wing in the velocity profile (see
Fig.~\ref{lya238}) implying a column density $N($\huno$)
\gsim 10^{19}$ \cm.
Unfortunately, the spectra of Q2139-4433 and Q2139-4434
do not cover the wavelength region where the
corresponding \huno\ \Lya\ lines should fall, while the
spectrum of Q2138-4427 does not cover 
that of the \cq\ doublet at this redshift. In 
the low resolution spectrum of Q2139-4434 by
\citet{fran:hew93}, an absorption line with equivalent
width $\sim 20$ \AA\ is present at this redshift, which
would correspond to a \Lya\ line with $N($\huno$) \sim
7\times 10^{19}$ \cm. 
We do not detect \cq\ absorption at this redshift in the
spectra of  Q2139-4433 and Q2139-4434 but we identify
neutral and singly ionised transition lines (\cdue,
\ouno, \sidue\ and \fed) with a simple
two-component velocity profile in Q2139-4434. 
Figure~\ref{hit238} shows two coincident transitions in
Q2138-4427 and Q2139-4434, they have a minimal velocity
separation of around 150 \kms, while the two LOSs are at
a transverse separation of $\sim 9\ h^{-1}$ Mpc. 


\begin{figure}
   \begin{center}
	 \resizebox{\hsize}{9cm}{\includegraphics{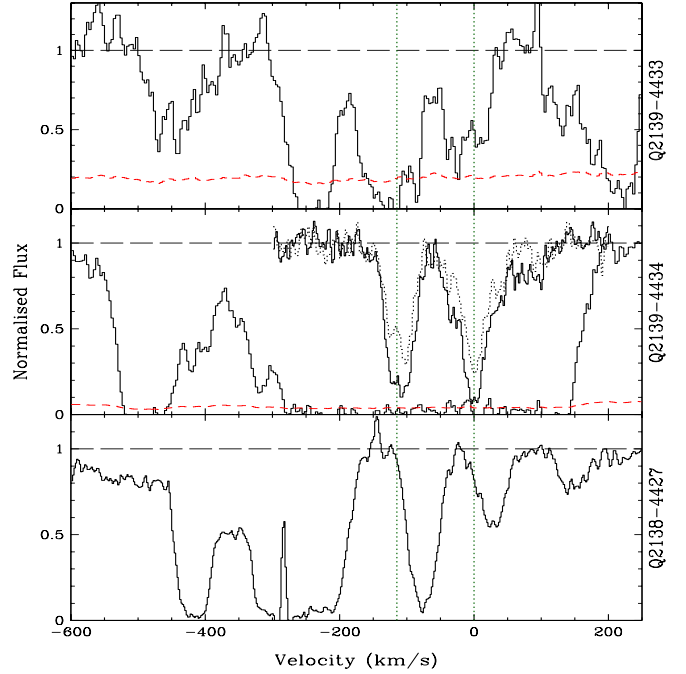}}
   \end{center}
   \caption{$z \sim 2.73$ - LLS at $z_{\rm a} \simeq
	 2.73557$ (origin of 
	 the velocity axes) in the spectrum of Q2139-4434
	 (middle panel). The \huno\ \Lya\ absorption line
	 is shown with superposed the lines of the \cq\
	 doublet. The other panels plot the corresponding
	 \huno\ \Lya\ line in the spectrum of  Q2139-4433
	 (top) and of Q2138-4427 (bottom). No metal lines
	 are associated to the 
	 latter two hydrogen absorptions. At this
	 redshift, the transverse spatial separation
	 between the LOSs to Q2139-4434 and Q2139-4433
	 is $\sim 1\ h^{-1}$ Mpc, between Q2139-4434 and
	 Q2138-4427 is $\sim 9\ h^{-1}$ Mpc and
	 between Q2139-4433 and Q2138-4427 is $\sim 8\
	 h^{-1}$ Mpc}   
\label{hit273}
\end{figure}

\vskip 12pt
\noindent
8) $z \simeq 2.73$ -
The system at $z_{\rm a} \simeq 2.73557$ in the spectrum
of Q2139-4434 is again a candidate LLS on the ground of
the observed ionic transitions. No metal lines are
detected within $\sim 3000$ \kms\ of this absorption
redshift along the LOS of Q2139-4433 and of Q2138-4427. 
On the other hand, the velocity profile of the observed
\huno\ \Lya\ absorptions follows that of the
\cq\ absorption associated to the LLS (see
Fig.~\ref{hit273}). 
Unfortunately, we cannot disentagle the velocity
structure of the LLS \Lya\ absorption since our spectrum
does not extend to the region where the higher lines in
the Lyman series are located. 

%

\begin{figure}
   \begin{center}
	 \resizebox{\hsize}{9cm}{\includegraphics{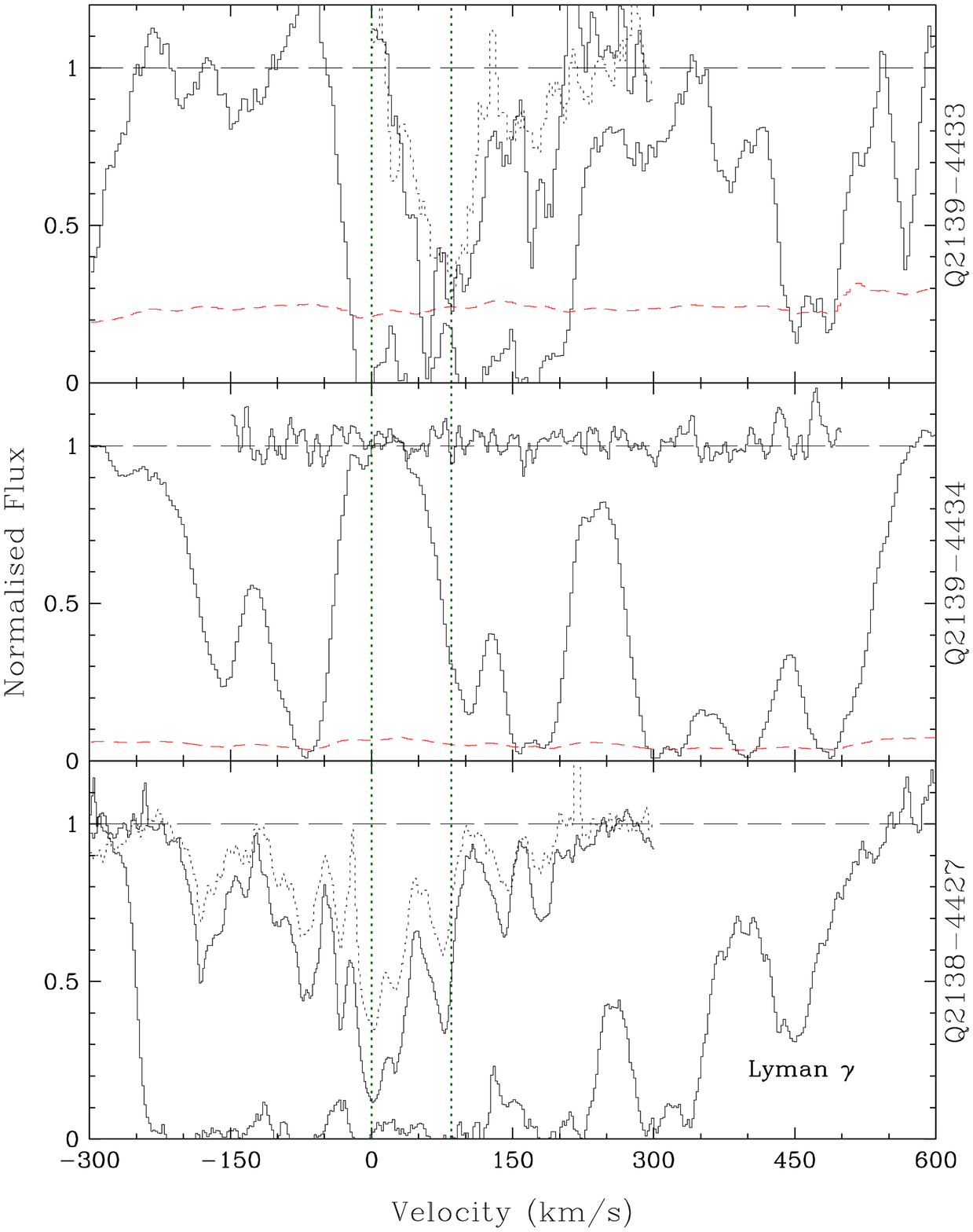}}
   \end{center}
   \caption{$z \sim 2.85$ - Coincident absorption systems
	 in the spectra of Q2139-4433 (top panel),
	 Q2139-4434 (middle panel) and Q2138-4427  
	 (bottom panel). The \huno\ \Lya\ transitions
	 are shown in the top and mid row and the 
	 \huno\ Lyman-$\gamma$ corresponding to the DLAS
	 in the spectrum of Q2138-4427 is in the bottom
	 row. Overplotted are the
	 corresponding \cq\ doublets, no metal absorption
	 is observed in the spectrum of Q2139-4434. The dotted
	 vertical lines mark the position of the main
	 components in \cq\ at $z_{\rm a} = 2.85153$ (origin of
	 velocity axes) and at $z_{\rm a} = 2.85262$. The
	 transverse spatial separations between the three
	 LOSs are about the same as those reported in the
	 caption of Fig.~\ref{hit273}}   
\label{hit285}
\end{figure}

\vskip 12pt
\noindent
9) $z \simeq 2.85$ - 
The DLAS at $z_{\rm a} \simeq 2.85$ in the
spectrum of Q2138-4427 coincides with a complex \huno\
\Lya\ absorption in the spectrum of Q2139-4434, with no
detectable associated metal transitions. 
On the other hand, we identify a saturated
\Lya\ absorption ($W_{\rm r} \simeq 1$ \AA) and a \cq\
doublet at  $z_{\rm a} \simeq 2.85262$  in the spectrum
of Q2139-4433 (see Fig.~\ref{hit285}), partially
superposing in redshift upon the \cq\ absorption
associated to the DLAS.  
The transverse spatial separation between the two LOSs at
this redshift is $\sim 9\ h^{-1}$ Mpc. 

This correlation could be interpreted as due to a gaseous
structure perpendicular to the LOSs and extending over
several Mpc in the direction defined by the three quasars.

\begin{table*}
\begin{center}
\caption{Summary of observed coincidences}\label{tab:hits}
\begin{tabular}{lclccccc}
\hline
&&&&&&& \\
Objects & Ident. & Redshift$^{\rm a}$ & $\Delta s^{\rm b}$ &
$\Delta v_{\rm min}^{\rm c}$ & log $N$(\huno) & $W_{\rm
r}(\lambda1548)$ & log $N$(\fed) \\  
& & & ($h^{-1}$ Mpc) & (\kms) & & \AA & \\
\hline 
&1 & 1.7874  & 0.87 &          & 13.8 & out & $<$~11.8 \\
&  & 1.78865 & & $>$~3000 & 19.0 & out & 14.5 \\ 
UM680 & 2 & 2.0352  & 0.92 & 300 & $>$~18 & 0.4$^{\rm d}$ & 12.8 \\ 
UM681 &   & 2.03215 &      &     & $>$~18 & 0.7$^{\rm d}$ & 13.4 \\
&3 & 2.12312 & 0.94 & 100 (\sidue) & $>$~17.3 & 0.5  & $<$~12.7 \\ 
&  & 2.12209 &      &              & $>$~17.3 & 0.44 & $<$~12.6 \\
\hline
&4 & 2.17115 & 5 & $>$~3000 & $>$~17.3 &  0.34 & 13.1 \\
&  & 2.167   &   &          &  &  $<$~0.01 & $<$~11.8 \\
Q2343+1232 & 5 & 2.43125 & 5.3 & 110 (\siq) & 20.35 & 1.1$^{\rm e}$ & 14.7 \\
Q2344+1228 &   & 2.4271  &     &     & $>$~15.9 & 0.7$^{\rm e}$ & $<$~12.5 \\
&6 & 2.549$^f$ & 5.3 &  & & &  \\
&  & 2.53788   &     &  & 20.4 & out & 14.1 \\
\hline
& 7 & out    & 1   &            &        & $<$~0.014 & $<$~12.3\\
& & 2.37977 & 9   & 150 (\fed) & 20  & $<$~0.008 & 13.4 \\
& & 2.38279 & 8 &              & $> 19$  & out & e.w. 1.2$^{\rm h}$  \\	
Q2139-4433 & 8 & 2.73258 &  &          & 16.5 & $<$~0.03 & $<$~12.9 \\
Q2139-4434 &   & 2.73557 &  & $>$~3000 & $> 17.3$ & 0.6 & 13 \\
Q2138-4427$^{\rm g}$ &   & 2.7323  &  &   &   &$<$~0.005& $<$~11.9 \\
&9  & 2.85262 &  & 0 (\cq) & & 0.5 & $<$~13 \\
&   & 2.85378 &  &         & 14.8 & $<$~0.007 & $<$~12.6 \\
&   & 2.85153 &  &         & 20.9  & 0.8 & e.w. 0.2$^{\rm h}$ \\
\hline
\end{tabular}
\end{center}
\scriptsize{
$^{\rm a}$ The reported redshifts correspond to the main
component of the associated metal absorption, if present;
or to the strongest \huno\ \Lya\ absorption closer to the
redshift of the high density system \\   
$^{\rm b}$ Transverse spatial separation between
the lines of sight; in the case of the triplet it refers
to the distance to the following object in the list \\
$^{\rm c}$ Minimal velocity separation between metal absorption
lines of the same ionic species in the coupled lines of
sight \\ 
$^{\rm d}$ \cq\ $\lambda\ 1548$ rest equivalent width from
\citet{sha:rob83} \\
$^{\rm e}$ \cq\ $\lambda\ 1548$ rest equivalent width from
\citet{sarg87} \\
$^{\rm f}$ Emission redshift of the paired QSO \\
$^{\rm g}$ Precise column density determination for the
metal lines in the spectrum of Q2138-4427 will be
reported by Ledoux et al. (in preparation) \\
$^{\rm h}$ Rest equivalent width in \AA
}
\end{table*}

\section{Discussion}

The expected number of DLAS ($N($\huno$) > 2\times
10^{20}$ \cm) and LLS ($2 \times 10^{17} < N($\huno$) <
2\times 10^{20}$ \cm) in the  
redshift interval covered by our 7 spectra as 
computed from their number density as a function of
redshift - $N(z)_{\rm DLAS} \simeq 0.055(1+z)^{1.11}$,
$N(z)_{\rm LLS} \simeq  0.27(1+z)^{1.55}$ \citep{slw00}-
is of 1 and 9, respectively. 
We detect 3 DLASs and 8 LLSs indicating that our lines of
sight are not strongly biased toward an overabundance of
high column density systems. 

The investigation of the nearby lines of sight at
the redshift of each of the previous systems, gives the
following results. 
Of the three DLASs: 2 coincide with metal systems with
\cq\ rest equivalent width $W_{\rm r}(\lambda1548)>
0.5$~\AA, and 1 is at less than 1000 \kms\ from the
emission redshift of the paired QSO, which in turn is
marking the presence of a high matter density peak
\citep[see][]{ellison01}.  
The transverse spatial separation over which these
coincidences happen varies between $\sim 5$ and 9
$h^{-1}$  Mpc. 

\begin{figure}
   \begin{center}
	 \resizebox{8cm}{8cm}{\includegraphics{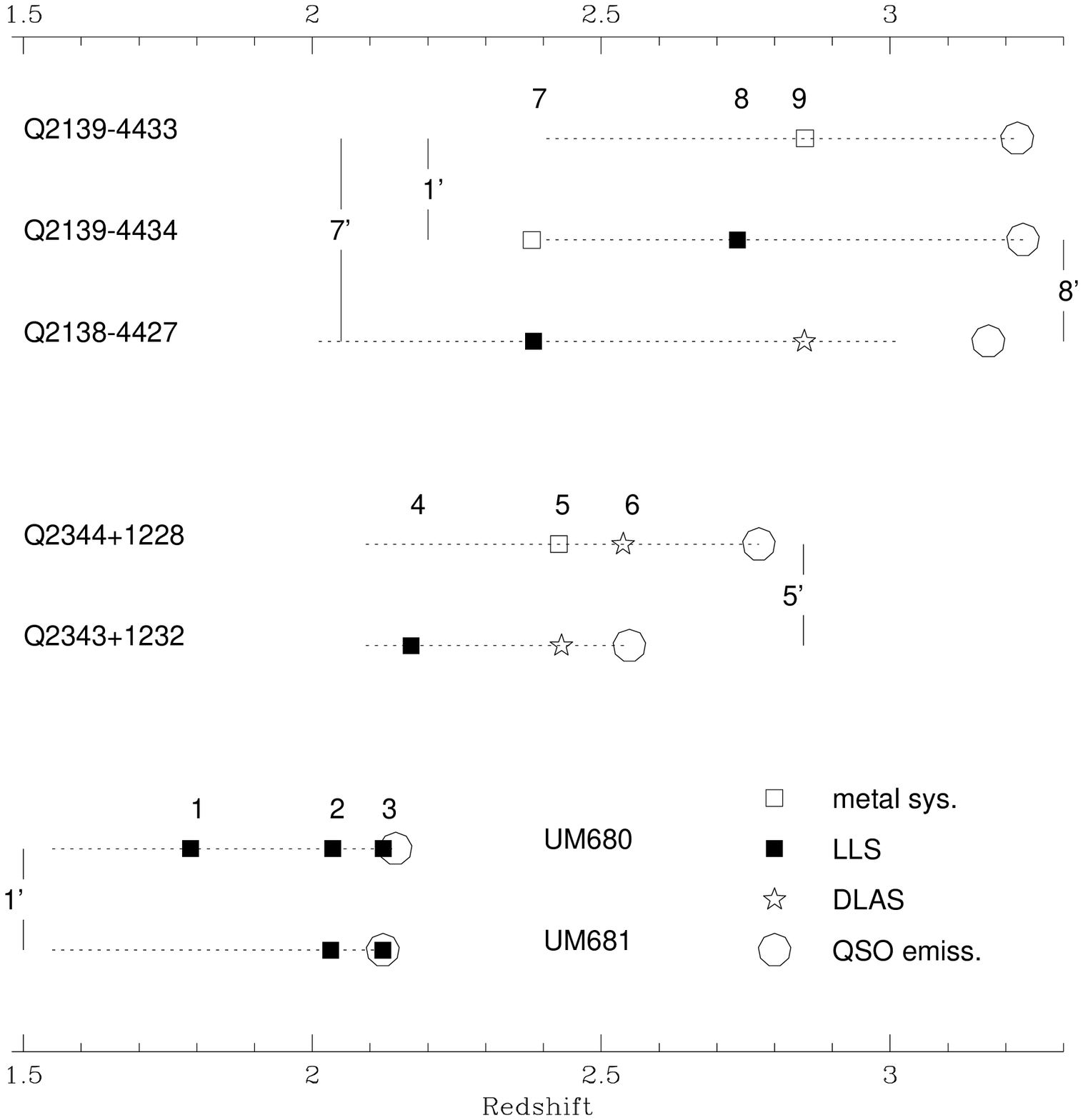}}
   \end{center}
   \caption{Summary of the observed coincidences as a
	 function of redshift. The dotted lines mark the
	 redshift range of the observed \Lya\ forests. 
         The angular separations of the quasars are
	 reported between the solid vertical lines. The
	 symbols are: open square for metal systems,
	 solid square for LLS with $2\times 10^{17} <
	 N({\rm HI})< 2\times 10^{20}$ \cm\ and star for
	 DLAS with $N({\rm HI})> 2\times 10^{20}$
	 \cm. The big open circles mark the 
	 emission redshift of the quasars}    
\label{hits}
\end{figure}  

As for the 8 LLSs: 4 of them form two
coinciding pairs at $z_{\rm a} \sim 2.03$ and 2.12 in the 
spectra of UM680 and UM681, their transverse spatial
separations are $\sim 920$ and 940 $h^{-1}$ kpc,
respectively. The LLS at $z_{\rm a} \sim 2.38$ in the 
spectrum of Q2138-4427 shows a coinciding metal system in
the spectrum of Q2139-4434 at
a transverse spatial separation $\sim 9\ h^{-1}$
Mpc. However, only low-ionisation transitions are
observed and no \cq. Furthermore, the \huno\ \Lya\ of the
latter system is outside our spectral range .    
The remaining 3 Lyman limit systems have corresponding
\Lya\ absorptions without associated metals within 3000
\kms.  

In summary,  we measure a
coincidence within 1000 \kms\ between high density
systems, in 5 cases out of 10.  
We exclude the coincidence at $z \sim 2.38$ in the
triplet, since it was not possible to determine the
\huno\ column density of the metal system. 
Figure~\ref{hits} shows a pictorial description of the
observed coincidences as a function of redshift; while in
Table~\ref{tab:hits} we report the main properties of the
matching absorption systems. 

In order to approximately compute the significance of our 
result, we consider the number density of \cq\ systems
with rest equivalent width $W_{\rm r} > 0.3$ \AA\ as a
function of redshift \citep{steidel90}.  The chance 
probability (in the hypothesis of null clustering) to
detect a \cq\ absorption line within 1000 \kms, 
between $z =2$ and 3, is ${\cal P}_{\rm exp} \simeq
0.004$.    
If we assume that a binomial random process rules the
detection or the non-detection of a coincidence, the
{\sl a posteriori} probability in the studied case is
$< 2.5 \times  10^{-10}$.   
The clustering signal is indeed highly
significant. 

Going back to our sample, the two coincidences 
in the spectra of UM680, UM681 at $\sim 1\ h^{-1}$ Mpc 
are closely related to the emitting  quasars. 
As recently claimed for associated absorption lines
\citep[e.g.][]{srppj00,kool01,hamann01}, the observed
absorption systems could arise in gas expelled
by a galactic ``superwind'' in a luminous starburst
associated with the formation of the quasar itself.  
Superwinds contain cool dense clouds which justify the
presence of low ionisation lines, embedded in a hot
($\sim 10^7$ K) X-ray-emitting plasma \citep[see][ and
references therein]{heckman96}. In low redshift galaxies,
outflow velocities of $10^2-10^3$ \kms\ and column
densities $N(H) \sim {\rm few}\ \times 10^{21}$ \cm\ have
been measured which are consistent with the observed
values.      

The remaining three coincident systems involve DLASs and 
are characterized by larger QSO pair separations. 
Damped systems at high redshifts are thought to
arise in large disks (e.g. Wolfe 1995) or in
multiple protogalactic clumps (Haehnelt et al. 1998;
Ledoux et al. 1998; McDonald \& Miralda-Escud\'e 1999).  
In either case they trace high matter density peaks and
they are possibly associated with Lyman-break galaxies
(M\o ller et al. 2002).  
The representation of these kind of objects in
hydrodynamical simulations \citep[e.g.][]{virgo98,cen98} 
shows that they lie in knots of $\sim 1\ h^{-1}$ Mpc
scale from which filaments several Mpc in length depart
in a spider-like structure. Star formation takes place in
the central condensation but also in some denser blobs
of matter along the filaments. 
The correlation on large scales observed around the DLAS
in our sample 
finds a likely explanation in this scenario \citep[see the
discussion in][]{francis01b}. 
For comparison, Lyman break galaxies at $z \sim 3$, which
are thought to have masses $M \sim 10^{11}$ M$_{\odot}$, 
show correlation lengths $r_0 \sim 2\ h^{-1}$  Mpc
\citep{giav98,cris01,arnouts02}.  

\section{Conclusions}

We have analysed new, high resolution UVES spectra of
two QSO pairs and a QSO triplet (refer to Table~\ref{obs}
and Sect.~2) focussing mainly on the clustering properties 
of high matter density peaks, traced by LLS ($2\times
10^{17} < N$(\huno)/\cm$ < 2\times 10^{20}$) and DLAS
($N$(\huno$) > 2\times 10^{20}$ \cm). 
The observed number of DLAS and LLS in the considered
lines of sight is in good agreement with the expected 
value.
 
\noindent
The relevant conclusions are the following: 

\begin{description}
\item[1.] in 4 cases out of 10 there is a metal system with
   \cq\ rest equivalent width $W_{\rm r} > 0.5$ \AA, in
   the paired line of sight within 1000 \kms\ of the
   redshift of the considered high column density
   absorption system. In 1 case, a DLAS matches the
   emitting QSO in the paired line of sight;

\item[2.] the correlation signal is highly significant in
spite of the small sample. The gas giving rise to the
coincidences close to the emission redshift of the QSOs
(\# 2 and 3 in Fig.~\ref{hits}) could be due to a
starburst-driven galactic superwind. In the other
cases (\# 5, 6 and 9), involving DLASs, the gas could be
in coherent filamentary or sheet-like structures of
several Mpc, the possible ancestors of present-day rich 
clusters;   

\item[3.] we measure the chemical abundance ratios in two 
DLAS and a sub-DLAS. In particular, we estimate the
ratios C/Fe, O/Fe and Si/Fe for the DLAS at $z\sim
2.53788$ in the spectrum of Q2344+1228. 
The abundance ratios of these elements are
consistent with solar values or very small enhancement at
a variance with what is 
observed for halo stars at the same metallicity in the
Milky Way. 
\end{description}

\begin{acknowledgements}
V.D. is supported by a Marie Curie individual fellowship
from the European Commission under the programme
``Improving Human Research Potential and the
Socio-Economic Knowledge Base'' (Contract
no. HPMF-CT-1999-00029). This work was supported in part
by the European Community RTN network ``The Physics of
the Intergalactic Medium''. It is a pleasure to thank
C. Ledoux for the UVES spectrum of Q2138-4427.   
\end{acknowledgements}

\bibliography{aamnem99,myref}
\bibliographystyle{apj}

\end{document}